Title: **Demonstration of a broadband-RF VLBI system at 16 Gbps data rate per station**

Author list:
Alan R. Whitney[1], <awhitney@haystack.mit.edu
Christopher J. Beaudoin[1], <cbeaudoin@haystack.mit.edu>
Roger J. Cappallo[1], <rjc@haystack.mit.edu>
Brian E. Corey[1], <bec@haystack.mit.edu>
Geoffrey B. Crew[1], <gbc@haystack.mit.edu>
Shepherd S. Doeleman[1], <dole@haystack.mit.edu>
David E. Lapsley[2], <dl99@me.com>
Alan A. Hinton[3], <a_hinton@comcast.net>
Stephen R. McWhirter[1], <russmc@haystack.mit.edu>
Arthur E. Niell[1], <aniell@haystack.mit.edu>
Alan E. E. Rogers[1], <aeer@haystack.mit.edu>
Chester A. Ruszczyk[1], <chester@haystack.mit.edu>
Daniel L. Smythe[1], <dls@haystack.mit.edu>
Jason SooHoo[1], <jsoohoo@haystack.mit.edu>
Michael A. Titus[1], <mat@haystack.mit.edu>

[1]MIT Haystack Observatory, Off Route 40, Westford, MA 01886
[2]OpenX Technologies, Inc., 20 E Del Mar Blvd., Pasadena, CA 91105
[3]Consultant, Mont Vernon, 21 Old Milford Rd, Mont Vernon, NH 03057

Abstract

The recent development of a relatively inexpensive 16-Gbps data-recording system based on commercial off-the-shelf technology and open-source software, along with parallel development in broadband Very Long Baseline Interferometry (VLBI) techniques, is enabling dramatically improved sensitivity for both astronomical and geodetic VLBI. The system is described, including the results of a demonstration VLBI experiment that illustrates a number of cutting-edge technologies that can be deployed in the near future to significantly enhance the power of the VLBI technique.

Keywords

Astronomical Instrumentation
Astronomical Techniques

Introduction

The technique of Very Long Baseline Interferometry (VLBI) makes use of many globally distributed radio telescopes both to obtain high resolution images of distant celestial objects and to measure the shape of the Earth and its orientation in space. Because the sources of radio emission are so weak, extremely sensitive receiving systems are required. This sensitivity can be improved by increasing the size of the radio antennas, reducing the noise contribution from the receivers, or, for continuum-frequency observations, increasing the recorded data rate. For existing antennas or when contemplating the economics of new antenna systems, greater continuum sensitivity is often best achieved by increasing the observing bandwidth, which translates directly into higher data rate. In this paper we report a single-baseline dual-polarization VLBI experiment with a 2-GHz aggregate-bandwidth in which data were recorded at 16-Gbits/sec at each station. This demonstration illustrates a number of cutting-edge technologies that can be deployed in the near future to enhance the sensitivity of astronomical and geodetic VLBI observing arrays.

In contrast to the current and recent generations of recorders used for VLBI, which are based on proprietary hardware components and controlling firmware, the new Mark 6 system uses commercial-off-the-shelf (COTS) technology and open source software. The high data rate is achieved by adopting modern high-performance motherboards and components writing simultaneously to 32 conventional magnetic disks.

Applications

*Astronomy*

As an astronomical imaging technique, VLBI is unrivaled for providing high angular resolution, as well for providing a means to study high brightness temperature cosmic phenomena with exceptional detail. At 2cm wavelength, inter-continental baselines (5000km) achieve resolutions (λ/D; λ is observing wavelength and D is projected baseline length) of 0.8 milliarcsec, and at 1.3mm wavelength on similar baselines, a resolution of 50 microarcsec can be obtained. The sensitivity of VLBI arrays is proportional to $B^{1/2}$ (B is total recorded bandwidth), and historically the bandwidth limit has been set by the capability of VLBI backends and recorders, not by radio-telescope receivers, which typically have bandwidths of several GHz. By contrast, the data capture rate of VLBI recorders was for many years limited by magnetic tape technology to 512Mb/s (Whitney 1996), which corresponds to 128 MHz of bandwidth (Nyquist sampled, 2-bits/sample). The adoption of hard-disk storage media and industry-standard hardware has begun to successfully address this mismatch in bandwidth between the telescope front end and VLBI instrumentation (Whitney et al 2010), enabling dramatic sensitivity improvements in VLBI networks including the Very Long Baseline Array (Walker et al 2007). These efforts have led to advances in astrophysical research that require both high angular resolution and high sensitivity, including faint afterglows of Gamma Ray Bursts (Pihlström et al 2007), searches for Active Galactic Nuclei (AGN) in starburst galaxies (Alexandroff et al 2012), and investigation of the Radio Quiet population of AGN (Bontempi et al 2012).

For VLBI observations at the shortest wavelengths (≤ 1.3mm) continued expansion of recorded bandwidth is particularly useful and critical for several important science objectives. Because of surface accuracy requirements, radio telescopes are typically smaller at these short wavelengths than at cm wavelengths, and increased bandwidth compensates for the smaller collecting area to some extent. In addition, atmospheric turbulence limits the VLBI integration time at short wavelengths to ≲ 10 sec, so that recording a wide bandwidth is essential in order to allow detection of faint sources (Rogers et al 1995). State of the art mm and submm receivers have bandwidths of ≳ 4 GHz, requiring VLBI recorders that can reach sustained data rates of order 16Gb/s in order to achieve maximum sensitivity.

Of particular scientific interest in this short wavelength regime are VLBI studies of nearby supermassive black holes, for which emission near the event horizon can be resolved and imaged. For SgrA*, the 4 million solar mass black hole at the Milky Way center (Reid 2009; Genzel et al 2010), the event horizon subtends an angle of 10 micro arcsec, presenting us with the opportunity to observe General Relativistic (GR) effects in a strong gravity regime. Recent 1.3mm VLBI observations have detected structures on scales of a few Schwarzschild radii towards SgrA*, which can be interpreted in the context of strong gravitational lensing at the event horizon (Doeleman et al 2008, Fish et al 2011, Broderick et al 2009). For M87, the giant elliptical galaxy whose core harbors a ~6.2 billion solar mass black hole, 1.3mm VLBI has been used to detect a similar sized structure (~5 Schwarzschild radii) at the base of the relativistic jet produced at the galaxy's core. The size of this structure, which is on the same scale as the Innermost Stable Circular Orbit (ISCO) of the black hole, has now set limits on the black hole spin and orbital direction of the accretion disk (Doeleman et al 2012). Extension of this work towards true imaging of these sources, which would allow tests of GR (Johannsen & Psaltis, 2010; Johannsen et al 2012), requires the higher bandwidth systems described herein.

*Geodesy*

Geodesy is the study of the shape and rotation of the Earth and their changes with time. VLBI provides the most accurate measurements of the shape of the Earth on a global scale. Twice-weekly VLBI observations using a global network coordinated by the International VLBI Service for Geodesy and Astrometry (Schlüter & Behrend 2007), in combination with the Satellite Laser Ranging network and the Global Navigation Satellite System network of antennas, provide the International Terrestrial Reference Frame (ITRF) to an accuracy of a few centimeters or better, with goals in the few-mm range for the next-generation VLBI2010 (Petrachenko et al 2009) system currently under development. The ITRF has become the basis for the civil surveying infrastructure in most countries, replacing less accurate country-wide or continent-wide reference systems that on occasion led to embarrassing inconsistencies, such as bridges that did not meet at boundaries. The IVS observations are used to determine a) UT1 minus UTC, which allows the maintenance of civil time at the accuracy needed by GPS for greatest accuracy, b) the International Celestial Reference Frame, composed of the positions of several thousand extragalactic radio sources, and c) nutation and precession of the Earth's spin axis, which tells us much about the structure of the Earth's core (Koot et al 2010).

For the geodetic VLBI system several hundred extragalactic broadband radio sources are needed, distributed as uniformly around the sky as possible (Schuh & Behrend, 2012). To move rapidly among the sources fast slewing antennas are needed, but this requirement for speed encourages smaller antennas for economic reasons. Therefore to make up for the lower sensitivity resulting from the smaller collecting area a higher data rate is needed to obtain the same sensitivity. To increase the uniformity of the source distribution generally requires making use of the weakest among the candidate sources, which also encourages adoption of the highest possible data rate.

While the antenna area, receiver temperature, and data rate determine the sensitivity per unit time, the sensitivity per "observation" can generally be increased by integrating for a longer time. However, for both applications that have provided the driver for the development of the new Mark 6 data-acquisition system, the integration time is limited. For astronomical observations made at frequencies of several hundred GHz, atmosphere turbulence limits coherent integrations to only a few tens of seconds or less. For the geodetic applications, studies made in preparing the specifications for the VLBI2010 system demonstrated that an observing rate of one or two scans per minute is desirable, which, with the move time required to get from source to source, provides a limit of less than ten seconds per observation for a fully developed system.

High level system description:

The observations reported here were made in 2012 June using the 12m antenna at Goddard Geophysical and Astronomical Observatory, Goddard Space Flight Center, Maryland, and the 18m Westford antenna at Haystack Observatory, Massachusetts; both antennas are part of the global geodetic VLBI network. The signal chain for each of these antennas (Figure 1) consists of a) a broadband dual linear polarization feed; b) a broadband Low Noise Amplifier (LNA) for each polarization; c) four-way splitters for each polarization; d) four frequency converters for translation of the signals from radiofrequency (RF) to intermediate frequency bands; e) two digital backend units (DBEs), and g) the Mark 6 16 Gbps digital recorder. The signal chain also includes a phase-calibration system which was not used in the experiments described in this paper.

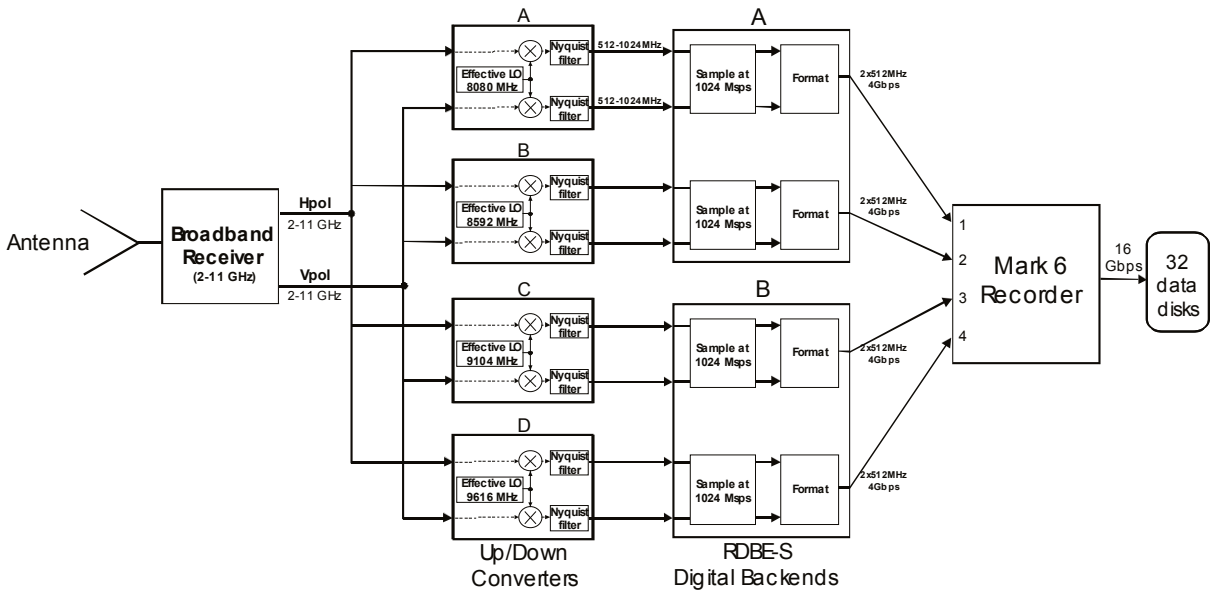

Figure 1: Simplified system level diagram of 4GHz bandwidth system recording to Mark 6 data recorder at 16Gbps

Except for the feeds, the receiving system is the same for each of the antennas. For the 12m antenna the feed is a Quadruple Ridged Flared Horn (Akgiray et al 2011), while for the Westford antenna an ETS-Lindgren 3164-05 quadridge feed was used. For both systems, the feed and LNAs are cooled to approximately 25K in a Dewar. The LNAs for both systems are Caltech 2-12 GHz single-ended MMICs. Figure 2 displays the typical system temperature performance of the receiver frontends installed on the Westford and GGAO antennas. Note that the system temperature stays below ~50K over this entire range except for some regions of radio interference near 2 GHz and 6 GHz.

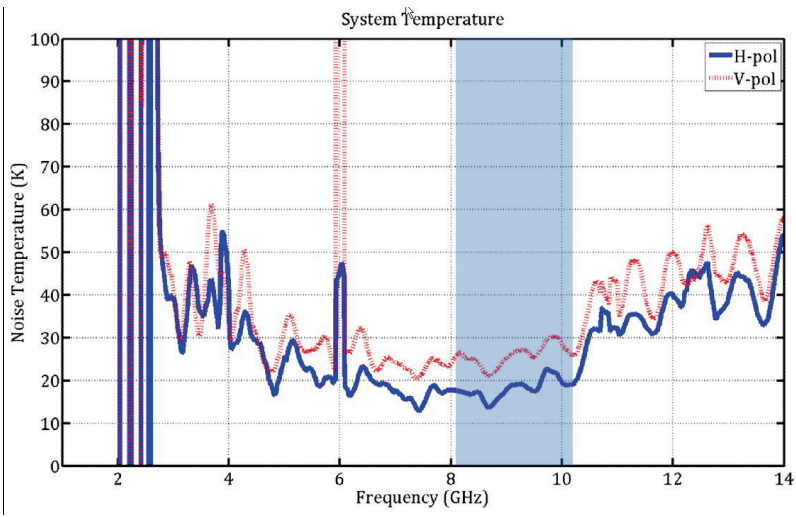

Figure 2: System temperature for horizontal and vertical polarizations as a function of frequency over the continuous band 2-14 GHz. The shaded area indicates the 2 GHz frequency range over which observations reported in this paper were conducted.

The RF for the two linear polarizations is brought to the control room on optical fibers. In the control room each polarization is split into four paths. A pair of polarizations is fed to each of four Up/Down Converters (UDCs) which translate both polarizations from four different RF bands to an IF in the range 0.5 GHz to 2.5 GHz. For these observations the four 512 MHz bands were selected to span the RF range 8592 MHz to 10640 MHz. Anti-aliasing filters in the output path of each UDC select a subset the IF range 512-1024 MHz corresponding to the second Nyquist zone of the digital back ends which follow.

Each DBE accepts the two polarizations from two bands, quantizes all four 512 MHz channels to two-bit samples, time-tags and formats the data into VLBI Data Interchange Format (VDIF) format (Whitney et al 2009), and outputs the formatted data to two separate 10-Gigabit Ethernet datastreams at 4 Gbps each, one for each band. The four Ethernet 4 Gbps streams (two from each DBE) are fed to the Mark 6 recorder through two dual-port 10-Gigabit Ethernet network interface cards.

Up/Down Converter

The up-down converter (UDC) translates an arbitrarily selected 2-GHz-wide slice of the RF spectrum to IF. The frequency translation is accomplished in two mixing stages (Figure 3), with an initial upconversion about a tunable LO frequency followed by a downconversion about a fixed frequency. The output IF passband is defined by the bandpass filter between the two mixers. By varying the first LO frequency, the spectral slice of the RF signal translated to IF can be varied. The downconversion is net upper sideband.

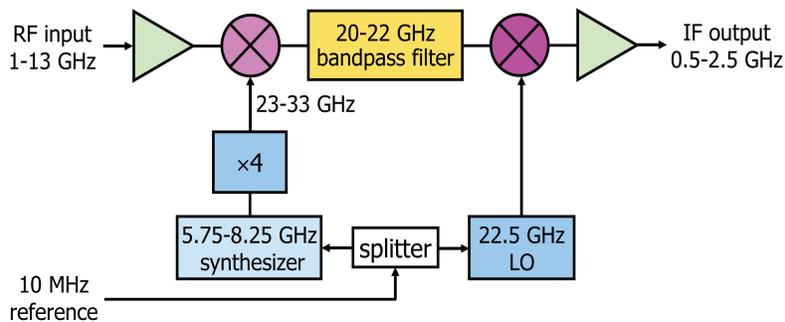

Figure 3: Simplified block diagram of up-down converter

The advantages of the UDC architecture include wide input and output frequency ranges, flexibility in tuning, and excellent image rejection. In the current implementation the rejection of the image and other unwanted responses is >70 dB. High rejection is particularly desirable when the RF input signal is broadband and has strong RFI at some frequencies.

Digital Backend

The wide-band single-channel Roach Digital Backend (RDBE-S) (see Figure 4) utilizes a Casper ROACH (Reconfigurable Open Architecture Computing Hardware) board (Parsons et al 2008) and supports two dual-input analog-to-digital converter (iADC) cards covering 2048MHz of spectrum. The FPGA personality accepts the four 512-Mhz analog channels from the four up/down converter, quantizes to 2 bits/sample, and outputs an aggregate of 8Gbps of data through two 10-Gigabit Ethernet ports. The output data are encapsulated in a UDP/IP payload for either long haul transmission or transmission to a local storage repository. The RDBE-S utilizes a standard 1-Gigabit Ethernet interface for control and monitoring.

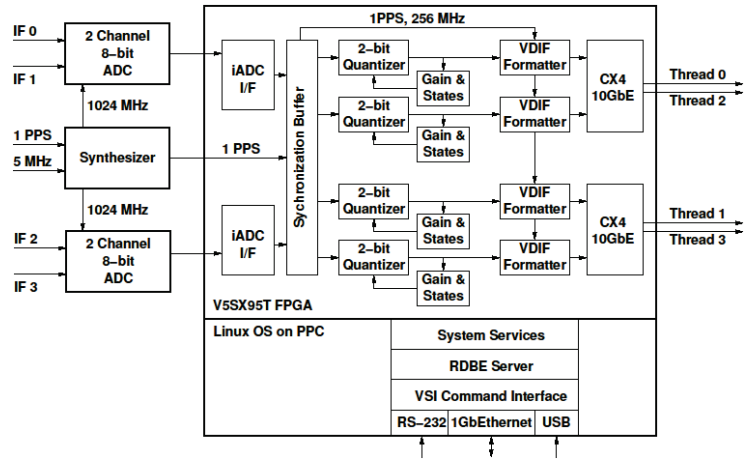

Figure 4: Block diagram of RDBE-S digital backend unit; large outline encloses FPGA-based ROACH board. Four 512MHz-bandwidth input IFs are converted into four 4Gbps VDIF-format data threads.

A separate FPGA personality (RDBE-H) can be used to configure the same ROACH hardware to process each of two independent 512MHz-wide analog channels through a polyphase-filter-bank to produce 32MHz-bandwidth channels more suited to geodetic-VLBI requirements.

The Mark 6 recorder

The Mark 6 VLBI data system is a joint collaborative development effort between MIT Haystack Observatory, NASA/GSFC High-End Network Computing group, and the Conduant Corporation, and is designed to support a sustained 16 Gbps data rate written to an array of disks. The hardware components were carefully selected for high performance and for compatibility with other system elements and use only COTS technology. The system operates under a fully open source Debian Linux distribution with the application software (also open source) written primarily in C++/C for the data plane and Python for the command and control interface.

The input to the Mark 6 can accommodate up to four 10GigE data ports, each operating independently at up to ~7Gbps with a maximum aggregate rate of ~16Gbps. The system supports SATA-interface disks packaged into modules (8 disks per module) that are inserted into the Mark 6 chassis and connected to the disk controllers via COTS external SATA cables. Each external SATA cable supports four disks, so that each module requires the connection of two such cables. Depending on required recording data rate, different numbers of simultaneously-operating disk modules are required. A single 8-disk module with modern disks will support 4Gbps of continuous recording; two modules (16 disks) are required for 8Gbps, 4 modules (32 disks) are required for 16Gbps. A prototype Mark 6 system is shown in Figure 5.

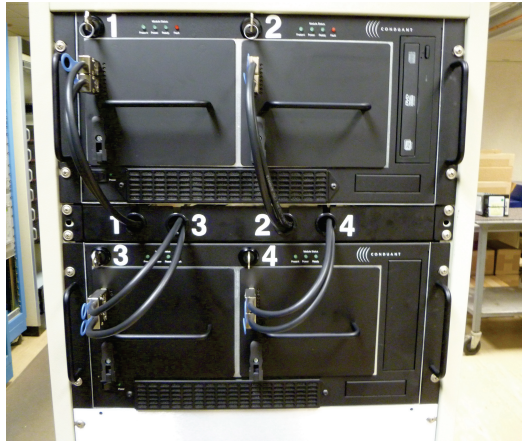

Figure 5: Photograph of prototype Mark 6 system.

VLBI has always pushed the technology of digital recording to the highest possible data rates, primarily due to the fact that, for the majority of VLBI observations, the achievable signal-to-noise ratio of a given system rises as the square root of the recording bandwidth. Figure 6a shows the evolution of VLBI recording capability from the origins of VLBI in the late 1960s through the present. Over this period the record data-rate capability has increased by more than four orders of magnitude from less than 1 Mbps in 1967 (magnetic tape) to 16 Gbps today (magnetic disks). At the same time, as shown in Figure 6b, the cost per Gbps of capability has dropped by almost five orders of magnitude and moved from highly proprietary tape-based systems to semi-proprietary disk-based systems (Mark 5 series), with the current 16-Gbps Mark 6 system utilizing almost fully COTS hardware with specially developed open-source software.

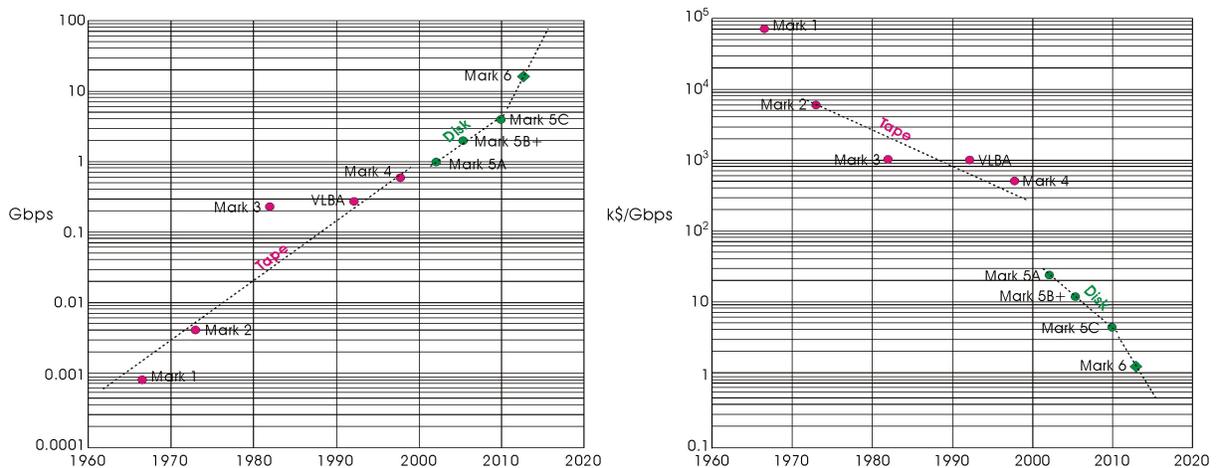

Figure 6: a) Evolution of VLBI recording-rate capability from 1967 to 2012, progressing more than four orders-of-magnitude from original 1967 magnetic tape to modern magnetic disks, b) parallel evolution of cost of recording in k$ per Gbps, which has dropped by almost five orders-of-magnitude during the same period.

The Observations

Several well-known sources of differing correlated flux density were observed with the dual-polarization (horizontal/vertical) receiver system at each site. The UDC frequencies were set to span the RF band from ~8GHz to 10GHz (see Figure 2) in four adjacent 512MHz bands. The programmable UDC attenuators were set manually to obtain approximately the proper level required for the samplers in the DBEs; a second level of gain adjustment, done for each scan, was applied digitally in the RDBE-S to achieve close-to-nominal sample-state statistics. Data recording was done using the Mark 6 system via four 4-Gbps Ethernet datastreams with Ethernet packet size of 8192 data bytes, plus 16 bytes of identifying header, in standard VDIF format; the aggregate 16-Gbps datastream at each station (2-GHz bandwidth in each of two polarizations) was recorded to four Mark 6 disk modules, each populated by eight standard 3.5-inch magnetic disks.

Observation sessions were scheduled for 17-18 May 2012 and 19 June 2012. Unfortunately, the May set of observations suffered from an equipment failure at the GGAO station that limited good data to be produced from only six of the planned eight 512MHz channels. The June observations were cut short after less than an hour by an emergency need to put Westford back into normal geodetic-VLBI service, so the only observation suitable for processing was on 3C84, the first source on the schedule.

After completion of the observations the data from both telescopes were electronically transferred to Haystack Observatory for processing.

Correlation and Fringe-Fitting

The data were cross-correlated on a 'DiFX' software-based correlation system (Deller et al 2011), processing horizontal-to-horizontal and vertical-to-vertical polarizations between the two stations. Four passes through the correlator, one for each IF band, were required to process the complete dataset.

Following correlation, the outputs from all four passes were transformed into a format compatible with the Haystack-developed *fourfit* fringe-fitting software. This fringe-fitting processing was unconventional and challenging due to the very large bandwidth and to the requirement to coherently combine the constituent frequency bands in both polarizations. Relative phase adjustments between correlated bands were determined manually in order to coherently combine the two bands in each polarization, then again manually to combine the two polarizations. In the future, this phase adjustment will be done automatically by measuring the phase of weak injected phase-calibration tones into the receiver.

Results

Figures 7 & 8 show the results of a 16 Gbps/station observation of 3C84 of 10 seconds duration taken on 19 June 2012. Although there was no ability to do careful calibration of the system, the correlation amplitude and signal-to-noise-ratio of the scan (Figure 7) are within the expected range. The time-segmented channel-by-channel amplitude and residual-phase data (Figure 8) also appear nominal.

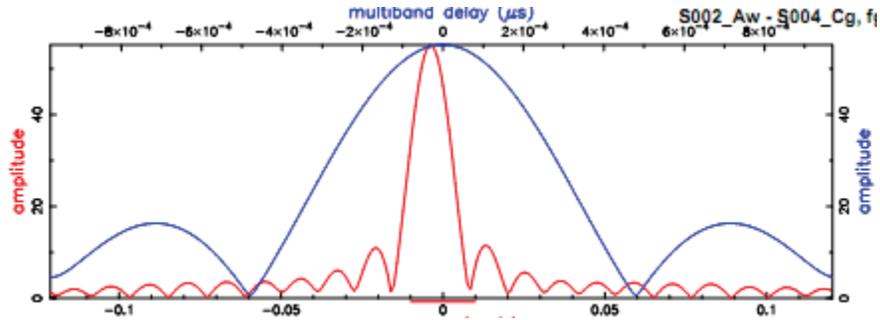

Figure 7: Correlation amplitude vs. multi-band residual delay (blue, scale at top) from 19 June 2012 observation of 3C84 at 16Gbps/station; red plot shows correlation amplitude as function of residual delay rate in nanoseconds/sec (scale at bottom). Correlation amplitude is ~5.5x10$^{-3}$ with a signal-to-noise ratio of ~940; HH and VV cross-correlations for all four bands (channels) were combined coherently for this result.

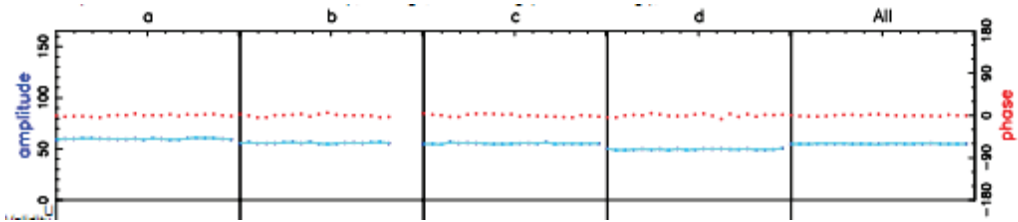

Figure 8: Plot of correlation amplitude (blue) and residual phase (red) vs. time for each of the four 512-MHz-bandwidth channels ('a' through 'd') and vector sum ('All') over the 10-second duration of the observation.

Although the results of the 17-18 May 2012 observations were somewhat compromised by the failure of one of the up-down converters, 16Gbps were recorded for a full 60 seconds on the Mark 6 system at each station, though only 12 Gbps of the recorded 16 Gbps were processed to obtain fringes on a weak (~0.2 Jy) source 0550+356 (Figures 9 & 10); as in the 3C84 observation above, only HH and VV correlations were done. Manual channel-to-channel adjustment phases for this source were obtained from observations of the nearby brighter source 0552+398 (7 Jy at 5 GHz).

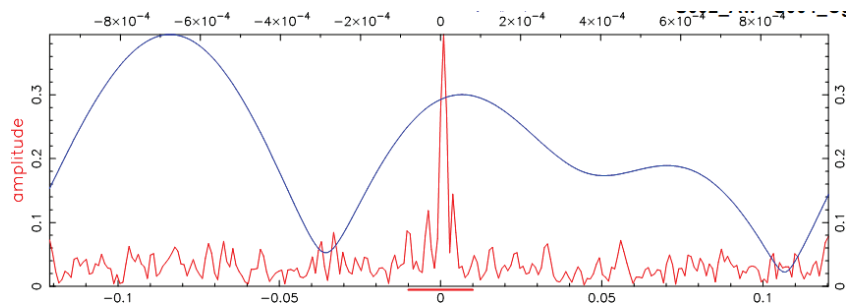

Figure 9: Correlation amplitude vs. synthesized multi-band residual delay (blue, scale at top in microseconds) from 17 May 2012 60-second observation of 0550+356; red plot shows correlation amplitude as function of residual delay rate in nanoseconds/sec (scale at bottom). Correlation amplitude is ~4.0x10$^{-5}$ with a signal-to-noise ratio of ~14; HH and VV cross-correlation for the three good bands (channels) were combined coherently for this result.

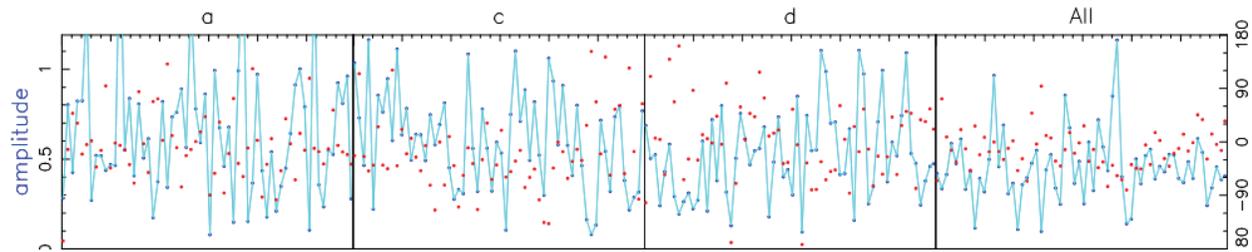

Figure 10: Plot of correlation amplitude (blue) and residual phase (red) vs. time for each for each of the three 512-MHz-bandwidth channels ('a', 'c', 'd') and vector sum ('All') over the 60-second duration of the observation.

Prospects, plans and future work

The 16 Gigagit/sec system used in these experiments, as well as some components of the broadband receiver system, are still in a process of maturation, with work continuing in the following areas:

1. The Mark 6 prototype system used in these experiments is an early RAID0 implementation and is susceptible to 'slow' disks impeding the sustainable record rate. For some applications, such as mm-VLBI or geodetic VLBI that tend to use sub-30-second scans, the RAM buffers supply sufficient capacity in the case that the sustained disk-writing rate is lower than the RAM capture rate, but may not work well for longer sustained observations. Work is currently in progress to develop a custom file system that is relatively tolerant of one or a few disks not operating fully to speed, allowing the system to sustain a higher throughput rate under these conditions.

2. A phase-calibration system that injects very short weak pulses into the broadband receiver has been developed to calibrate the relative phases of the individually-correlated frequency bands so they can be coherently combined without manual intervention.

Acknowledgements

Modifications of the control/monitor software to support these observations were done by Ed Himwich of NVI Inc.